\documentclass[10pt,here]{article}

\usepackage[all]{xy}
\usepackage{amsmath,amssymb}
\usepackage{graphicx}

\usepackage[all]{xy}
\usepackage{epstopdf}

\begin{document}

\begin{center}
\scshape
CENTRE DE PHYSIQUE TH\'EORIQUE \footnote{\, Unit\'e Mixe de
Recherche (UMR) 6207 du CNRS et des Universit\'es Aix-Marseille 1 et 2 \\ \indent \quad \, Sud Toulon-Var, Laboratoire affili\'e \`a la 
FRUMAM (FR 2291)} \\ CNRS--Luminy, Case 907\\ 13288 Marseille Cedex 9\\
FRANCE
\end{center}
\vspace{1cm}
\begin{center}
{\huge\bfseries 
Quasi-quantum groups from strings}
\end{center}
\vspace{0.5cm}
\begin{center}
{\it \large Talk given at the conference "Noncommutative Geometry and Physics"}
\end{center}
\vspace{0.5cm}
\centerline{\it \large Orsay, April 23-27, 2007}
\vspace{1cm}
\begin{center}
{\large Jan-H. Jureit\footnote{\, jureit@cpt.univ-mrs.fr}${}^,$\footnote{\, and University Kiel, Germany},
\quad Thomas Krajewski\footnote{\, krajew@cpt.univ-mrs.fr} }
\vspace{1cm}
\end{center}

{\bfseries\scshape Abstract:} 

%\section*(Abstract)

Motivated by string theory on the orbifold ${\cal M}/G$ in presence of a Kalb-Ramond field strength
$H$, we define the operators that lift the group action to the twisted sectors. These operators turn out to generate the
quasi-quantum group $D_{\omega}[G]$, introduced in the context of conformal field theory by R. Dijkgraaf, V. Pasquier and P. Roche \cite{dijkgraaf}, with $\omega$ a 3-cocycle determined by a series of cohomological equations in a tricomplex combining de Rham, \u{C}ech and group cohomologies. We further illustrate some properties of the quasi-quantum group from a string theoretical point of view. This work is based on \cite{paper}, from which a full-fledged treatment may be extracted. 

\vspace{1cm}

%\tableofcontents    

\vspace{4cm}
%\vskip 1truecm
%PACS-93:\\
%\indent MSC-91: 
%\vskip 1truecm
\noindent June 2007\\
\noindent
CPT-P31-2007

\newpage

 \section{Introduction}

We present a path integral derivation of the operators $T$ that lift the action of a finite group $G$ to the twisted sectors of bosonic strings on the orbifold ${\cal M}/G$ in a 3-form magnetic background $H$. For example, this applies to orbifolds of the WZW models with ${\cal M}=\mbox{SU(N)}$, $H=\frac{k}{12\pi}\mbox{Tr}(g^{-1}dg)^{3}$ and $G$ a finite subgroup of $\mbox{SU(N)}$. However, we do not treat this example in detail here and stick to some general algebraic aspects of the operators $T$ that can be deduced from the geometry of string worldsheets.

We proceed in complete analogy with the motion of a particle in a 2-form magnetic field $B$ that provides a projective representation of $G$ on the particle's wave functions. Because processes involving strings are highly constrained by the
geometry, the algebra generated by the operators $T$ has a richer structure which  turns out to be the quasi quantum group $D_{\omega}[G]$, introduced in the context of conformal field theory by R. Dijkgraaf, V. Pasquier and P. Roche \cite{dijkgraaf}. $D_{\omega}[G]$ is a quasi-triangular quasi-Hopf algebra whose product is determined by the commutation with propagation
\begin{equation}
T\quad\parbox{1.5cm}{\mbox{\includegraphics[width=1.5cm]{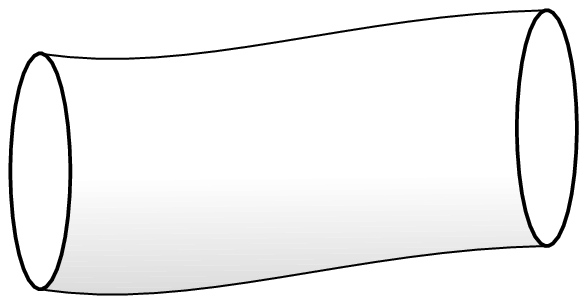}}}
\quad=\quad
\parbox{1.5cm}{\mbox{\includegraphics[width=1.5cm]{./pics/_minicy.eps}}}
\quad T,
\end{equation}
and whose coproduct follows from the commutation with the most basic interaction
\begin{equation}
T\quad
\parbox{1.5cm}{\mbox{\includegraphics[width=1.5cm]{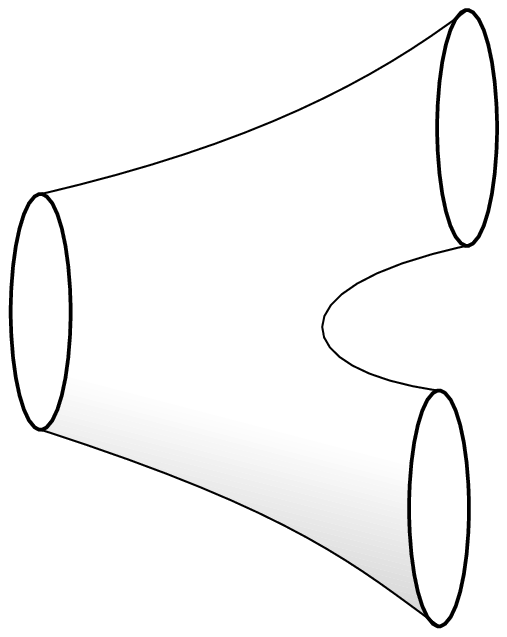}}}
\quad=\quad
\parbox{1.5cm}{\mbox{\includegraphics[width=1.5cm]{./pics/_minipant.eps}}}
\quad\Delta T.
\end{equation}

To carry out the geometric analysis in the path integral formalism, we adapt to the twisted sectors
the techniques developed by K. Gaw\c{e}dzki for topological actions \cite{gawedzki} and view string states as wave functionals $\Psi(X)$ of the string's embedding in the space-time manifold ${\cal M}$.

\section{Magnetic translations for a particle in a $B$-field}

%\underline{\bf Magnetic amplitude for a particle}

%\smallskip

For a particle moving on a manifold ${\cal M}$ in an external magnetic field $B$ represented as a closed 2-form with integral periods, the wave functions are sections of a line bundle ${\cal L}$ over ${\cal M}$ 
with a connection $\nabla$ of curvature $B$. The Hilbert space ${\cal H}$ of the theory is made out of square integrable wave functions and a large class of classical observables may be quantized as operators on ${\cal H}$. The most familiar example in physics is the motion of a charged particle in presence of Dirac's magnetic monopole, whose wave functions can be understood as sections of a line bundle over the two dimensional sphere. In this case, the integrality condition amounts to the quantization of the product of the electric and magnetic charges.  

In the path integral approach, the kernel of the evolution operator is represented as a sum over paths,
\begin{equation}
{\cal K}(y,x)=\mathop{\int[D\varphi]}
\limits_{\begin{subarray}{c}\varphi(a)=x\cr\varphi(b)=y\end{subarray}}\,
\mathrm{e}^{-S[\varphi]}\,{\cal A}[\varphi],\label{feynmanpart}
\end{equation}
\noindent
with $S$ the kinetic part of the classical action. The effect of the magnetic field is encoded in the magnetic amplitude ${\cal A}[\varphi]$ which is an extra  phase factor that weights the various paths. It can interpreted geometrically as the holonomy of the connection $\nabla$, whereas ${\cal K}(y,x)$ is a map from the the fiber at $x$ to the fiber at $y$ of the line bundle ${\cal L}$. 

Both ${\cal L}$ and ${\cal A}[\varphi]$ can be constructed using a good open cover $\left\{U_{i}\right\}$ of ${\cal M}$. Starting with the local expression of the globally defined magnetic field $B_{i}=B\vert_{U_{i}}$, one derives locally defined real 1-forms $A_{i}$ and $U(1)$ valued functions by solving the following system of equations,
\begin{equation} 
\left\{
\begin{array}{lcrcl}
B_{i}&=&\mbox{d} A_{i}\quad&\mbox{on}&\quad U_{i},\cr
A_{j}-A_{i}&=&\mathrm{i}\,\mbox{d}\log f_{ij} \quad&\mbox{on}&\quad
U_{i}\cap U_{j},\cr f_{jk}(f_{ik})^{-1}f_{ij}&=&1
\quad&\mbox{on}&\quad U_{i}\cap U_{j}\cap U_{k},\label{bundle}
\end{array}
\right.
\end{equation}
Using nowhere vanishing local sections, global sections of ${\cal L}$ may be identified with complex valued functions $\psi_{i}$ defined on $U_{i}$ that fulfill $\psi_{i}=f_{ij}\psi_{j}$ on $U_{i}\cap U_{j}$ and the connection is given by the covariant derivative $\nabla\psi_{i}=d\psi_{i}-\mathrm{i}A_{i}\psi_{i}$.

Using a triangulation of the path by edges and vertices such that each of its elements  is included in a single open set of the cover, the magnetic amplitude (holonomy of the
connection along the path) reads
\begin{equation}
% -----------------------
{\cal A}_{ij}[\varphi]=\exp \mathrm{i}\left\{\sum_{l_{\alpha}\in
l}\int_{l_{\alpha}} \varphi^{\ast}A_{i_{\alpha}}\right\}\,
\mathop{\prod}_{\begin{subarray}{c} l_{\alpha}\in l\\ v_{\beta}\in
\partial l_{\alpha}
\end{subarray}}
f_{i_{\alpha}j_{\beta}}^{-\epsilon_{\alpha\beta}}(\varphi(v_{\beta})),
\label{holo} \end{equation} where $\epsilon_{\alpha\beta}=+1$ if
$l_{\alpha}$ is arriving at $v_{\beta}$ and $-1$ if it is leaving.

%
%%%%%%%%%%%%%%%%%%%%%%%
% Picture of the path %
\begin{figure}[h]\centering
\includegraphics[width=7cm]{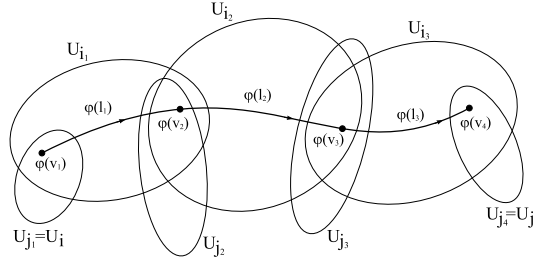}
\caption{Triangulation of a path and associated open sets}
\end{figure}
%%%%%%%%%%%%%%%%%%%%%%%
%
Using the equations \eqref{bundle}, it is easy to show that the magnetic amplitude is independent of all the arbitrary choices (triangulation, assignements of the open sets) that we made, except on the boundary, in accordance with its interpretation as the  holonomy of $\nabla$. Besides, the construction is also invariant  under gauge transformations given by $U(1)$ valued functions $f_{i}$ on the open sets $U_{i}$ that transform $A_{i}$, $\psi_{i}$ and $f_{ij}$.

%\medskip

%\underline{\bf Projective group action on wave functions}

%\smallskip

In a classical theory, a symmetry may be defined as an action\footnote{To get a left action on differential forms we choose a right action on $X$} of a group $G$ on the manifold  ${\cal M}$ that leaves the dynamics of the particle invariant. In our setting, this means that the kinetic part of the action $S$ is genuinely invariant and that $g^{\ast}B=B$.

At the quantum level, any $g\in G$ should be promoted to a unitary  operator $T_{g}$ acting on the Hilbert space ${\cal H}$ and commuting with the evolution operator. In the geometric language, the action of $T_{g}$ on the wave functions should read
\begin{equation}
T_{g}\psi(x)=\phi_{g}(x)\,\psi(x\!\cdot\!g),\label{Tpart}
\end{equation}
which may be decomposed into a pull-back action on the section (taking a section of ${\cal L}$ to a section of $g^{\ast}{\cal L}$) followed by a phase multiplication (realizing the equivalence between ${\cal L}$ and $g^{\ast}{\cal L}$).

The phases are determined in the path integral formalism by the commutation of $T_{g}$ with propagation $KT_{g}=T_{g}K$. Because the kinetic term is genuinely invariant, the commutation relation amounts to
\begin{equation}
{\cal A}[\varphi\!\cdot\!g]= \phi_{g}^{-1}(y)\,{\cal
A}[\varphi]\,\phi_{g}(x)\label{gApart}.
\end{equation}
A local expression for the phases may be obtained in concrete situations by solving the equations
\begin{equation}
\left\{
\begin{array}{rclrl}
g^{\ast}A_{i}-A_{i}&=&\mathrm{i}\,\mbox{d}\log\phi_{g;\, i} &\mbox{on}& U_{i},\\
g^{\ast}f_{ij}\,(f_{ij})^{-1}&=&\phi_{g;\,j}\,(\phi_{g;\,i})^{-1}
&\mbox{on}& U_{i}\cap U_{j},
\end{array}
\right.
\label{phipart}
\end{equation}
which determine $\phi_{g;\, i}$ up to a constant phase.

As is often the case in quantum mechanics, these operators only form a projective representation of the symmetry group,
\begin{equation}
T_{g}T_{h}=\omega_{g,h}\,T_{gh},\label{projpart}
\end{equation}
with the group 2-cocycle
\begin{equation}
\omega_{g,h}=\phi_{h}(x\!\cdot\!g)\,\phi^{-1}_{gh}(x)\phi_{g}(x).
\end{equation}
Note that $\omega$ is a constant whereas $\phi_{g;\,i}$ are in general non constant, locally defined functions of $x$.

Let us now try to define an orbifold particle on the quotient ${\cal M}/G$. Physical states on the quotient are expected to be made out of invariant states in ${\cal H}$. Because states correspond to wave functions only up to a constant phase, the invariance condition of a wave function reads
\begin{equation}
T_{g}\psi=\alpha_{g}\psi,
\end{equation}
with $\alpha$ a $U(1)$ valued function on $G$. If we further apply $T_{h}$ on both sides of this equation and use $\eqref{projpart}$, we find that $\omega$ is necessarily a coboundary,
\begin{equation}
\omega_{h,g}=\alpha_{h}\alpha^{-1}_{hg}\alpha_{g}=(\delta\alpha)_{h,g},\label{triviality}
\end{equation}
with $\delta$ the group coboundary operator that we shall define in full generality in the next section. Accordingly, the cohomology class of $\omega$ is an obstruction to the
existence of a quantum theory on ${\cal M}/G$ or, in geometric terms, to the existence of a bundle on the quotient. 

These operators may be considered as generalizations of magnetic translations for a particle on  ${\Bbb R}^{N}$ in a uniform magnetic field $B=\frac{1}{2}\,B_{\mu\nu}\,dx^{\mu}\wedge dx^{\nu}$. Restoring the charge of the particle $e$ and Planck's constant, the magnetic translations read, in the symmetric gauge,
\begin{equation}
T_{a}\psi(x)=\mathrm{e}^{-\frac{\mathrm{i}e}{2\hbar}B_{\mu\nu}a^{\mu}x^{\nu}}\,\psi(x+a)
\label{torexplicit}
\end{equation}
with $a\in{\Bbb R}^{D}$. The magnetic translations obey the multiplication law
\begin{equation}
T_{a}T_{b}=\mathrm{e}^{\frac{\mathrm{i}e}{2\hbar}B_{\mu\nu}a^{\mu}b^{\nu}}\;T_{a+b},
\end{equation}
with a 2-cocycle whose phase is $\pi$ times the ratio of the magnetic flux through the parallelogram spanned by $a$ and $b$ to the flux quantum $\Phi_{0}=\frac{2\pi\hbar}{e}$.
 
It is important to notice that the motion on the torus ${\Bbb R}^{D}/{\Bbb Z}^{D}$ is well defined if and only if the cocycle appearing in the subrepresentation of ${\Bbb Z}^{D}\subset{\Bbb R}^{D}$ is trivial, since otherwise no invariant state exist. This condition is equivalent to the quantization of the flux through the unit cell of the lattice, expressed in natural units. 

\section{Higher gauge fields and their symmetries}

%\underline{\bf Magnetic fields for closed strings}

%\smallskip

Let us now consider a closed string moving on ${\cal M}$ in the presence of an external magnetic field. As the string evolves, it sweeps out a 2-dimensional surface $\Sigma$ so that the dimensions of all geometrical objects are naturally increased by one.  In particular, the magnetic potential is the Kalb-Ramond  2-form $B$ instead of the 1-form $A$ and is related to its 3-form fieldstrength by $H=dB$. The Kalb-Ramond field enters the world-sheet path integral as an extra phase factor, in complete analogy with the particle's case.

However, assuming that $H$ is globally exact is much too restrictive since this is not the case for the celebrated WZW models describing strings moving on the group manifold ${\cal M}=\mbox{SU}(N)$ with $H=\frac{k}{12\pi}\mbox{Tr}(g^{-1}dg)^{3}$. In the general case, given $H$ which we assume to be a closed 3-form with integral periods,  the potentials are only locally defined. They are obtained by solving the series of equations,
\begin{equation}
\left\{
\begin{array}{lcrcl}
H_{i}&=&\mbox{d} B_{i}\quad&\mbox{on}&\quad U_{i},\cr
B_{j}-B_{i}&=&\mbox{d} B_{ij}\quad&\mbox{on}&\quad U_{i}\cap
U_{j},\cr B_{jk}-B_{ik}+B_{ij}&=&\mathrm{i}\,\mbox{d}\log f_{ijk}
\quad&\mbox{on}&\quad U_{i}\cap U_{j}\cap U_{k},\cr
f_{jkl}(f_{ikl})^{-1}f_{ijl}(f_{ijk})^{-1}&=&1 \quad&\mbox{on}&\quad
U_{i}\cap U_{j}\cap U_{k}\cap U_{l}.\label{gerb}
\end{array}
\right.
\end{equation}
These equations admit two layers of gauge transformations: a gauge transformation of the potentials as well as a gauge transformation of the gauge transformation itself. On the geometrical side, these equations define a gerb with connection, which is a higher dimensional generalization of a line bundle with connection. Using a transgression map, the 3-form $H$ defines a 2-form on the loop space $L_{\cal M}$ and the gerb is associated to a line bundle with connection over $L_{\cal M}$, whose sections define the string wave functions. Then, the magnetic amplitude in the worldsheet path integral is naturally identified with the holonomy of the previous connection.

%\medskip

%\underline{\bf Tricomplex combining de Rham, Cech and group
%cohomologies}

%\smallskip

To handle the high number of locally defined fields entering the analysis of the invariance properties of the gerb, it is convenient to introduce a tricomplex combining de Rham, \u{C}ech and group cohomologies. Cochains $C_{p,q,r}$ are de Rham forms
of degree $p$, defined on $(q+1)$-fold intersections of a "good
invariant cover"\footnote{All open the sets are invariant and their intersections contractible, but they may not be connected. } $U_{i_{0}}\cap\dots\cap U_{i_{q}}$ and functions of $r$ group indices. It admits three commuting differentials:

\begin{itemize}

\item
the de Rham differential in the $p$ direction (with the convention that we always use $\mathrm{i}\,\mbox{d}\,\mbox{log}$ for functions);
\item
the \u{C}ech coboundary $\check\delta$ in the $q$ direction 
\begin{equation}
(\check\delta c)_{i_{0}\dots i_{q}}=
\sum_{k=0}^{q}(-1)^{k}c_{i_{0}\dots,\check{i}_{k}\dots,i_{q}}
\end{equation}
where $\check{i}_{k}$ means that the index $i_{k}$ has been
omitted;
\item
the group coboundary $\delta$ in the $r$ direction
\begin{equation}
(\delta c)_{g_{0},\dots,g_{r}}=g_{0}^{\ast}c_{g_{1},\dots, g_{r}}
+\sum_{k=1}^{r}(-1)^{k}\,c_{g_{0},\dots,g_{k-1}g_{k},\dots,g_{r}}
+(-1)^{r+1}c_{g_{0},\dots, g_{r-1}},
\end{equation}
where $g^{\ast}$ is the pullback action on forms.
\end{itemize}

%\begin{figure}[h]
%\[ \xymatrix @R=0.4pc @C=1.2pc {
%            & \bullet \ar[rr] & & \bullet \\
%            \bullet \ar[rr] \ar[ru] & & \bullet \ar[ru] & \\
%            & \bullet \ar'[r][rr] \ar'[u][uu] & & \bullet\ar[uu] \\
%            \bullet \ar[rr]_{\mathrm{d}} \ar[uu]^{\delta} \ar[ru]^{\check\delta} & &\bullet %\ar[uu] \ar[ru] &  }
%\]
%\caption{Three commuting differentials}
%\end{figure}

For any fixed value of $r$, we have a \u{C}ech-de Rham bicomplex,
\begin{equation}
 C_{r,s}^{\mathrm{tot}}=\bigoplus_{p+q=s}C_{p,q,r},
\end{equation}
equipped with the Deligne differential defined by ${\cal
D}=\mp\mbox{d}\pm\check{\delta}$ fulfilling ${\cal D}^{2}=0$ and
$\delta{\cal D}={\cal D}\delta$.

%\medskip

%\underline{\bf Symmetries of 2-form potentials}

%\smallskip

\begin{figure}[h]\centering
\includegraphics[width=6cm]{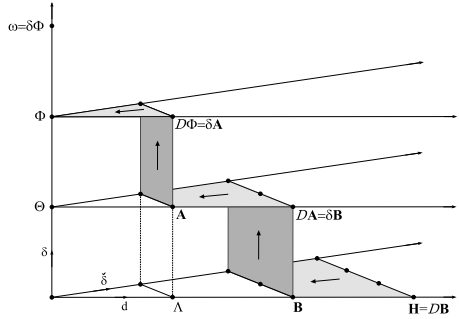}
 \label{_tricomplex}
\caption{Cohomological equations}
\end{figure}

Starting with a closed 3-form $H$ on ${\cal M}$ such that $g^{\ast}H=H$ for any $g\in G$, we first define ${\bf H}=(H_{i},0,0,1)\in C_{0,3}^{\mathrm{tot}}$. Because $H$ is closed and globally defined one has  ${\cal D}{\bf H}=0$ and the invariance of $H$ translates into $\delta {\bf H}=0$. Assuming that the cohomology of the total \u{C}ech-de Rham complex vanishes in degree 1 and 2, which holds for the group manifold of simply connected simple Lie Group like $\mbox{SU}(N)$, we solve a series of cohomological equations leading to a constant 3-cocycle $\omega$. First, we rewrite the defining equations of the gerb \eqref{gerb} as ${\bf H}={\cal D}{\bf B}$ with ${\bf B}=\left( B_{i},B_{ij},f_{ijk}\right)\in C_{0,2}^{\mathrm{tot}}$. Then, the invariance of $H$ leads to
\begin{equation}
\qquad {\cal D}\delta {\bf B}=\delta{\cal D}{\bf
B}=\delta{\bf H}=0 \quad\Rightarrow \quad \exists\,{\bf A}\in
C_{1,1}^{\mathrm{tot}}\quad\mbox{s.t.}\quad\delta {\bf B}={\cal
D}{\bf A}.
\end{equation}
Next, we reiterate the same reasoning starting with $\delta{\bf A}$
\begin{equation}
\qquad {\cal D}\delta {\bf A}=\delta{\cal D}{\bf
A}=\delta^{2}{\bf B}=0 \quad\Rightarrow \quad \exists\,{\bf \Phi}\in
C_{2,0}^{\mathrm{tot}}\quad\mbox{s.t.}\quad\delta {\bf A}={\cal
D}{\bf \Phi}.
\end{equation}
Finally, we recover the constant 3-cocycle from $\omega=\delta\Phi$. In the tricomplex, one inverts the Deligne operator to move towards the third axis and uses the group coboundary to move upwards, as illustrated in figure \ref{_tricomplex}.

%%%%%%%%%%%%%%%%%%%%%%%%%%%%%
% Picture of the tricomplex %
%%%%%%%%%%%%%%%%%%%%%%%%%%%%%
%

As expected, there are two layers of gauge ambiguities in the definition of ${\bf B}$. First one can transform ${\bf B}$ into ${\bf B}+{\cal D}\Lambda$ with $\Lambda\in C_{0,1}^{\mathrm{tot}}$. The gauge transformation of the gauge transformation sends $\Lambda$ to $\Lambda+{\cal D}\Xi$ with $\Xi\in C_{0,0}^{\mathrm{tot}}$ and leaves ${\bf B}$ invariant. There is an extra gauge ambiguity in the definition of ${\bf A}$ and $\Phi$ alone that replaces the latter by ${\bf A}+{\cal D}\Theta$ and $\Phi+\delta\Theta$ with $\Theta\in C_{1,0}^{\mathrm{tot}}$, which we shall refer to as a {\it secondary gauge transformation}.
Lastly, there is also a constant multiplicative ambiguity in the definition of $\Phi$ which changes $\omega$ by a coboundary.   

From a string theoretical viewpoint, the first gauge ambiguity will constrain the magnetic amplitude for a free propagating string, while secondary gauge transformations will be crucial in determining the interaction vertex. Finally the constant ambiguity, when satisfying the cocycle condition, is intimately tied with discrete torsion.

Since ${\bf A}$, $\Phi$ and $\omega$ are group cochains, it is helpful to evaluate them on elements of $G$, so that their defining relations read
\begin{equation}
\left\{
\begin{array}{rcl}
{\cal D} {\bf A}_{g}&=&g^{\ast}{\bf B}-{\bf B},\cr
{\cal D} \Phi_{g,h}&=&g^{\ast}{\bf A}_{h}-{\bf A}_{gh}+{\bf A}_{g},\cr
\omega_{g,h,k}&=&g^{\ast}\Phi_{h,k}\,\Phi^{-1}_{gh,k}\,\Phi_{g,hk}\,\Phi^{-1}_{g,h}.
\end{array}
\right.
\label{groupexplicit}
\end{equation}
As a convention, we always assume that all the group cochains are normalized, i.e. they vanish whenever one of their arguments equals the identity of the group. This is always possible by exploiting the gauge freedom. If we further display the \u{C}ech indices and assume $\omega=1$, we recover the equations displayed by E. Sharpe \cite{sharpe}. The equation $\omega=1$ is a further consistency condition for the orbifold to exist but cannot be derived solely form the invariance of $H$. At the geometrical level, $\omega$ is an obstruction to the existence of a push-forward of the gerb on ${\cal M}$ to a gerb on the quotient ${\cal M}/G$ \cite{newgawedzki}.

\section{String propagation and group action on twisted sectors}

%\underline{\bf Magnetic amplitude for twisted sectors}

%\smallskip

Recall that the orbifold procedure in string theory consists in two steps. First we add to the closed string Hilbert space the twisted sectors which are open strings on ${\cal  M}$ that close up to an element of $G$.  Loosely speaking, the Hilbert space ${\cal H}_{w}$ of twisted sectors of winding $w$ is made out of wave functions $\Psi$, whose arguments are maps $X:\,[0,2\pi]\rightarrow{\cal M}$ such that $X(2\pi)=X(0)\cdot w$. Then, we define the orbifold states as those vectors in the Hilbert space that are invariant under the action of $G$. This requires a lift of the action of $G$ on ${\cal H}_{w}$ that will be defiend in the sequel.

Let us now consider a free propagating string of winding $w$. The associated worldsheet $\Sigma$ is a cylinder and the effect of the magnetic field is encoded in a magnetic amplitude that has to be inserted in the path integral. To construct the magnetic amplitude one has to cut the cylinder to make it simply connected. Then, one integrates the local potentials contained in ${\bf B}$ (on the triangulated cylinder with a cut) and ${\bf A}_{w}$ (along the cut),
\begin{equation}
{\cal A}[\varphi]=\mbox{e}^{\mathrm{i}\int_{\Sigma}{\bf
B}+\mathrm{i}\int_{x}^{y}{\bf A}_{w}}.
\end{equation}
%%%%%%%%%%%%%%%%%%%%%%%%%%%%%
% Picture of the cylinder %
\begin{figure}[h]\centering
\includegraphics[width=7cm]{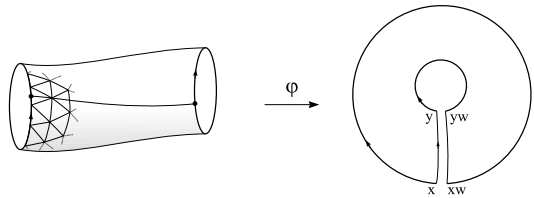}
\label{_cylinder}
\caption{Cylinder with triangulation and cut}
\end{figure}
%%%%%%%%%%%%%%%%%%%%%%%%%%%%%
%
The explicit expression of $\int_{\Sigma}{\bf B}$ involves integrals of the 2-forms of ${\bf B}$ on the triangles and of the 1-forms along the edges, as well as $U(1)$ valued functions evaluated on the vertices. Similarly, one integrates the 1-forms of ${\bf A}$ and evaluate the scalars for the associated triangulation of the cut. Because of lack of space, we are unable to display here the detailed construction and refer to \cite{paper} for a detailed account.

The rationale behind this construction is to first write for arbitrary twisted sectors the amplitude valid at zero winding involving only ${\bf B}$, as can be found in \cite{gawedzki}. Because we are dealing with open strings on ${\cal M}$, this amplitude suffers from several flaws that we correct by including a contribution of ${\bf A}_{w}$. This ensures, for instance, gauge invariance and invariance under homotopic changes of the cut.

%\medskip

%\underline{\bf Stringy magnetic translations and their algebra}

%\smallskip

As for a particle (see eq. \eqref{Tpart}), the stringy magnetic translations $T_{g}^{w}: {\cal
H}_{w^{g}}\rightarrow{\cal H}_{w}$ lifting the group action to the twisted sectors consist in
a pull-back action on the wave functions followed by a phase multiplication,

\begin{equation}
T_{g}^{w}\Psi_{w^{g}}(X)=
\Gamma_{w,g}(x)\,\mathrm{e}^{-\mathrm{i}\int_{x}^{xw}{\bf A}_{g}}\,
\Psi(X\!\cdot\!g), \label{definition}
\end{equation}
with $\Gamma_{w,g}=\Phi_{g,w^{g}}\Phi^{-1}_{w,g}$ and $w^{g}=g^{-1}wg$. The precise form of this factor is dictated by the requirement of the commutativity of the stringy magnetic translations with propagation. Assuming that all terms in the action but the magnetic amplitude are genuinely invariant, the expression of these phases follow from the comparison of the magnetic amplitude of a cylinder and of its lift by $g$, in complete analogy with eq. \eqref{gApart}.

As expected, stringy magnetic translations form a projective representation on the twisted sectors.
Indeed, composing two such operators we get,
\begin{equation}
T_{g}^{w}T^{v}_{h}=\delta_{v,w^{g}}\,\frac{\omega_{w,g,h}\,\omega_{g,h,w^{gh}}}{\omega_{g,w^{g},h}}\,
T_{gh}^{w},\label{product}
\end{equation}
which is identical to the multiplication law of the quasi-quantum group $D_{\omega}[G]$ \cite{dijkgraaf}. It is interesting to notice that the phase factor of the multiplication law admits a combinatorial interpretation in terms of tertrahedra representing the 3-cocycle. 
%
% Picture PRISM1
\begin{figure}[h]\centering
\includegraphics[width=5cm]{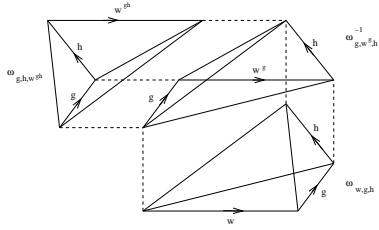}
 \label{_prism1}
\caption{Combinatorial interpretation of the product}
\end{figure}

%\bigskip

%\underline{\bf Orbifold and triviality of $\omega$}\\

%Let us end this section by a short comment on the orbifold procedure. First, one gathers all %the twisted sectors that belong to the same conjugacy class. This is necessary because the %action of $g$ takes a string of winding $w$ to a string of winding $w^{g}$. Next, on %identifies in each conjugacy class the states that are invariant up to a global phase. %Because $D_{\omega}[G]$ induces a projective representation on the conjugacy classes %\cite{maillard}, an argument similar to the one leading to \eqref{triviality} for indicates %that the phase factor of the multiplicaation law should be removable by a suitable change of %phase of $T_{g}^{w}$. This is indeed the case for the cohomologicqlly triviql cocycles %introduced in \cite{coste} (see also \cite{dvvv} for an early discussion).

\section{Quasiquantum group structure form string interactions}

%\underline{\bf Interactions}

%\smallskip

When compared to a theory of particles, the main feature of a theory of strings lies in the fact that their interactions are highly constrained by the geometry. Indeed, processes involving $m$ incoming strings that interact and produce $n$ outgoing ones are associated with surfaces whose boundaries are made of $m+n$ circles. Any of these surfaces can always be decomposed into cylinders and pair of pants, which reflects the fact that strings interact by joining and splitting.

The most basic interaction involves pair of pants whose magnetic amplitude contributes to the decay ${\cal
H}_{vw}\rightarrow {\cal H}_{v}\otimes{\cal H}_{w}$ of a string of winding $vw$ into a first string of winding $v$ and a second one of winding $w$. The magnetic amplitude for such a process is
\begin{equation}
 {\cal A}[\varphi]=
\mathrm{e}^{\mathrm{i}\int_{\Sigma}{\bf B} +\mathrm{i}\int_{x}^{t}{\bf
A}_{vw}+\mathrm{i}\int_{t}^{y}{\bf A}_{v}+\mathrm{i}\int_{tv}^{z}{\bf
A}_{w}} \;\Phi_{v,w}^{-1}(t),\label{pant}
\end{equation}
where we recognize three cylinder amplitudes for propagation of strings of windings $vw$, $v$ and $w$. Besides, one has to insert $\Phi_{v,w}$ at the splitting point to maintain invariance under secondary gauge transformations. 

%%%%%%%%%%%%%%%%%%%%%%%%%%%%%%%%%
% Picture of the Pant %
\begin{figure}[h]
\centering
\includegraphics[width=6cm]{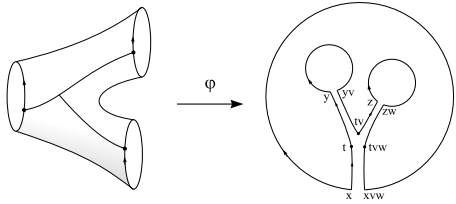}
 \label{_hose}
\caption{pair of pants}
\end{figure}
%%%%%%%%%%%%%%%%%%%%%%%%%%%%%%%%%
%

%\medskip

%\underline{\bf Derivation of the coproduct}

%\smallskip

The pair of pants has to be inserted in a path integral contributing to a decay process with an initial string in ${\cal H}_{vw}$ and final ones in ${\cal H}_{v}\otimes{\cal H}_{w}$. The requirement that stringy magentic translations commute with this process determines the action of $G$ on the tensor product of two twisted sectors. It turns out that there is an extra phase departing form the tensor product representation $T_{g}^{v}\otimes T_{h}^{w}$ on ${\cal H}_{v^{g}}\otimes{\cal H}_{w^{w}}$,  from which we read the coproduct,

\begin{equation}
\Delta(T_{g}^{u})=\mathop{\sum}\limits_{vw=u}\,\frac{\omega_{v,w,g}\,\omega_{g,v^{g},w^{g}}}{\omega_{v,g,w^{g}}}\,
T_{g}^{v}\otimes T_{g}^{w}.\label{coproduct}
\end{equation}

\begin{figure}[h]
\centering
\includegraphics[width=6cm]{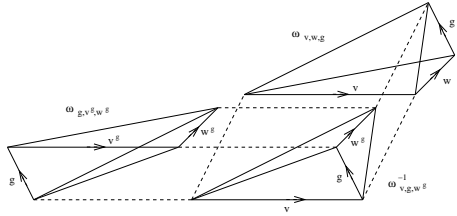}
\caption{Combinatorial interpretation of the coproduct}
\label{coprodpic}
\end{figure}

This extra phase also admits a combinatorial interpretation in terms of tetrathedra, as illustrated in figure \ref{coprodpic}.

The operators $T_{g}^{w}$ generate the quasi-quantum group $D_{\omega}[G]$ which is a quasi-triangular quasi-Hopf algebra deformation of the quantum double of the group algebra of $G$, induced by the magnetic background. Its representations form a quasi-tensor category,
which, can be thought as a generalization of the category of representations of groups. This is the reason why they provide a general framework that can encompass all possible symmetries we may encounter in physical theories. The general ideas leading to quasi-Hopf symmetry have been introduced by G. Mack and V. Schomerus \cite{schomerus}.

Let us now illustrate some of these points from a string theoretical point of view.

%\medskip

%\underline{\bf Quasi-tensor category}

%\smallskip

Roughly speaking, a quasi-tensor category is a category equipped with a tensor product and an operation that interchanges two factors in the tensor product. Usually the tensor product is associative only up to an isomorphism, the Drinfel'd associator, and the interchange of two factors is implemented through a representation of the braid group. We refer the interested reader to \cite{chari} for a thorough survey of quasi-tensor categories and content ourself with some examples illustrating how the associator and the braiding arise in our context.

%%%%%%%%%%%%%%%%%%%%%%%%%%%%%%%%%
% Picture of the 3-Pants %
\begin{figure}
\centering
\includegraphics[width=6cm]{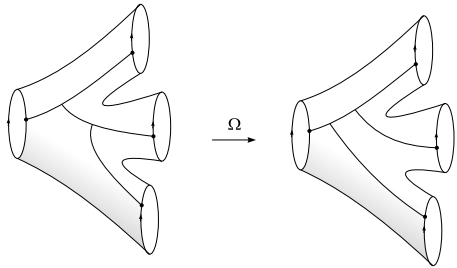}
\label{associator}
\caption{Different cuts leading to the associator}
\end{figure}
%%%%%%%%%%%%%%%%%%%%%%%%%%%%%%%%%

Let us consider the tree level decay of a single string of winding $uvw$ to three strings of windings $u$, $v$ and $w$. The corresponding worldsheet may be obtained by gluing two pair of pants in two different ways, as illustrated in figure \ref{associator}. The two worldsheets differ only by their cuts which induces a phase difference in the corresponding magnetic amplitudes. On the one side we have $\Phi^{-1}_{u,vw}u^{*}\Phi^{-1}_{v,w}$, while on the other side we have $\Phi^{-1}_{uv,w}\Phi^{-1}_{u,v}$ so that the two amplitudes differ by $\omega_{u,v,w}=(\delta\Phi)_{u,v,w}$. This is precisely the Drinfel'd associator
defined by
\begin{equation}
\Omega=\mathop{\sum}\limits_{u,v,w}\omega_{u,v,w}^{-1}\,
T_{e}^{u}\otimes T_{e}^{v}\otimes T_{e}^{w},
\end{equation}
measuring the failure of the coassociativity of the tensor product,
\begin{equation}
\left( \mathrm{id} \otimes \Delta \right)\circ \Delta=
\Omega\big(\left(\Delta\otimes\mathrm{id} \right)\circ\Delta\big)\Omega^{-1}.
\label{quasi666}
\end{equation}
On the physical side, this means that one process ends in $\left({\cal H}_{u}\otimes{\cal H}_{v}\right)\otimes {\cal H}_{w}$ while the other one ends in ${\cal H}_{u}\otimes\left({\cal H}_{v}\otimes {\cal H}_{w}\right)$. The two vectors only differ by the global phase $\omega_{u,v,w}$, so that they represent the same physics.

%%%%%%%%%%%%%%%%%%%%%%%%%%%%%%%%%
% Picture of the Braiding       %
\begin{figure}
\centering
\includegraphics[width=6cm]{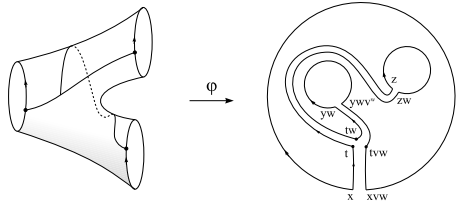}
\caption{Braiding}
\label{braiding}
\end{figure}
%%%%%%%%%%%%%%%%%%%%%%%%%%%%%%%%%
%

As a second illustration, let us consider the braiding. Because of the cut, the magnetic amplitude imposes an order to the outgoing strings so that the corresponding process ends in ${\cal H}_{v}\otimes{\cal H}_{w}$. Changing the cut as in figure \ref{braiding}, we end up in ${\cal H}_{w}\otimes{\cal H}_{v^{w}}$. This is implemented by a braid group action on the tensor product, whose generators are constructed by composing the flip with the $R$-matrix
\begin{equation} 
{\cal R}=\mathop{\sum}\limits_{v,w}
T_{e}^{v}\otimes T_{v}^{w}.
\label{Rmatrix}
\end{equation}
Of course, when more than two strings are involved, it is necessary to use the Drinfel'd associator before braiding, in order to make sure that the two strings being interchanged are not separated by parentheses. All this provides a representation of the braid group thanks to the quasi Yang-Baxter equation obeyed by ${\cal R}$ and $\Omega$.

The definition of a quasi-Hopf algebra also involves the antipode whose analog in the theory of group representation is the contragredient representation. At the string theoretical level, it implements the orientation reversing operation, allowing to switch between  the incoming and outgoing nature of the strings. 

Another important feature of quasi-Hopf algebra is the possibility of slightly changing the their structure by a Dirnfel'd twist. Such an operation is intimately related to discrete torsion in string theory.

%\underline{\bf Global anomalies and triviality of $\omega$

Finally, it is worthwhile to point out that the cocycle $\omega$ also appears when considering modular invariance of higher loop amplitudes. Indeed, one can easily write down the magnetic amplitude for a torus by gluing together the two ends of a cylinder. However, the latter fail to be modular invariant: acting with the generators $S$ and $T$ of the modular group of the torus changes the amplitudes by certain combinations of the 3-cocycle $\omega$. Thus, higher loop amplitudes seem to be flawed by global anomalies when $\omega$ is non trivial. Thus, the triviality of the 3-cocyle may be considered as a consistency condition for the orbifold in a magnetic background.

\section{Conclusion and outlooks}

With a view to the construction of an orbifold string theory, we have presented the general form of the operators that realize the action of a finite group $G$ on the twisted sectors in a magnetic 3-form background. The resulting operators generate the quasi-quantum group $D_{\omega}[G]$, which was originally introduced in the theory of holomorphic orbifolds \cite{dijkgraaf}. Our derivation is based on the construction of the magnetic amplitudes which encode the effect of 3-form background when inserted in the path integral and proceeds in complete analogy with the case of a particle in a magnetic field. The 3-cocycle is related to the 3-form $H$ by a series of cohomological equations and its triviality appears as a consistency condition for the orbifold string theory.

In a certain sense, it can be said that the quasi quantum group is a higher dimensional generalisation of the projective group representation, as summarized in the following table.

\renewcommand{\arraystretch}{1.5}

\bigskip

\centerline{\begin{tabular}{|c|c|} \hline
Particles&Strings\\
\hline
2-form $B$&3-form $H$\\
\hline
Line bundle&Gerb\\
\hline
2-cocycle $\omega$&3-cocycle $\omega$\\
\hline Projective representation&Quasi-quantum group\\
\hline
\end{tabular}}

\bigskip

As a higher dimensional generalization of projective group representations, quasi-quantum groups may play a significant role in the theory of higher dimensional extended objects. Indeed, following M. Douglas, discrete torsion for open strings
induces a projective group action on the branes \cite{douglas}. Thus, 
the orbifold group action for open membranes in the presence of the M-theory
discrete torsion, identified as a 3-cocycle $\alpha$ in \cite{sharpeMtheory}, may lead to an action of the  quasi-quantum group $D_{\alpha}[G]$ on the wave functions of the endlines of the membrane.

\section*{Acknoledgements}

T.K. would like to thank Philippe Roche for stimulating discusssions and the organizers of the workshop {\it "Noncommutative Geometry and Physics"} for  providing the opportunity to give a talk.

\end{document}